\documentclass[preprint,secnumarabic,tightenlines,amssymb,amsmath,nobibnotes,aps,nofootinbib,showpacs]{article}%
\usepackage[caption=false]{subfig}
\usepackage[T1,T5]{fontenc}
\usepackage[Gray,squaren,thinqspace,thinspace]{SIunits}
\usepackage{setspace}
\usepackage{color,braket,a4wide}
\usepackage{amsmath}
\usepackage{amsfonts}
\usepackage{verbatim}
\usepackage{amssymb}
\usepackage{epstopdf}
\usepackage{graphicx}
\usepackage{float}
\usepackage{hyperref}
\usepackage{color}%
\providecommand{\U}[1]{\protect\rule{.1in}{.1in}}
\definecolor{darkgreen}{rgb}{0,0.35,0}

\begin{document}

\title{\vspace{-2cm}\noindent\textbf{Double non-perturbative gluon exchange: an update on the soft Pomeron contribution to $pp$ scattering}}
\author{\textbf{F.~E.~Canfora$^{a}$}\thanks{%
fcanforat@gmail.com}\,\,, \textbf{D.~Dudal}$^{b,c}$\thanks{%
david.dudal@kuleuven.be}\,\,, \textbf{I.~F.~Justo}$^{d}$\thanks{%
igorfjusto@gmail.com}\,\,, \\\textbf{P.~Pais}$^{a,e}$\thanks{%
pais@cecs.cl}\,\,, \textbf{P.~Salgado-Rebolledo}$^{f}$\thanks{%
patricio.salgadot@uai.cl}\,\,, \textbf{L.~Rosa}$^{g,h}$\thanks{%
rosa@na.infn.it}\,\,,
\textbf{D.~Vercauteren}$^{i}$\thanks{vercauterendavid@dtu.edu.vn}\\
{\small \textnormal{$^{a}$ Centro de Estudios Cient\'{\i}ficos (CECS),
Casilla 1469, Valdivia, Chile}}\\
{\small \textnormal{$^{b}$ KU Leuven Kulak, Department of
Physics, }} \\
{\small \textnormal{\phantom{$^{b}$} Etienne Sabbelaan 53 bus 7657, 8500 Kortrijk, Belgium}}\\
{\small \textnormal{$^{c}$ Ghent University, Department of Physics and
Astronomy,}}\\
{\small \textnormal{\phantom{$^{c}$}  Krijgslaan 281-S9, 9000 Gent, Belgium}}\\
{\small \textnormal{$^{d}$ UFES - Universidade Federal do Esp\'{i}rito Santo,}}
\\
{\small \textnormal{\phantom{$^{d}$}Departamento de Qu\'{i}mica e F\'{i}sica, Centro de Ci\^{e}ncias Exatas, Naturais e da Sa\'{u}de,}}
\\
{\small \textnormal{\phantom{$^{d}$} Alto Universit\'{a}rio, s/n 29500-000, Guararema, Alegre, ES, Brasil }}\\
{\small \textnormal{$^{e}$ Physique Th\'{e}orique et Math\'{e}matique, Universit\'{e} Libre de Bruxelles}} \\
{\small \textnormal{\phantom{$^{e}$} and International Solvay Institutes, Campus Plaine C.P.~231, B-1050 Bruxelles, Belgium}}\\
{\small \textnormal{$^{f}$ Facultad de Ingenier\'ia y Ciencias and UAI Physics Center, Universidad Adolfo Ib\'a\~nez,}}\\
{\small \textnormal{Avda. Diagonal las Torres 2640, Pe\~nalol\'en, Santiago, Chile.}}\\
{\small \textnormal{$^{g}$ Dipartimento di Fisica, Universit\'{a} di Napoli
Federico II, Complesso Universitario di Monte S.~Angelo,}}\\
{\small \textnormal{\phantom{$^{f}$} Via Cintia Edificio 6, 80126 Napoli,
Italia}}\\
{\small \textnormal{$^{h}$ INFN, Sezione di Napoli, Complesso Universitario
di Monte S.~Angelo,}}\\
{\small \textnormal{\phantom{$^{g}$} Via Cintia Edificio 6, 80126 Napoli, Italia}}\\
{\small \textnormal{$^{i}$ Duy T\^an University, Institute of Research and
Development,}}\\{\small \textnormal{\phantom{$^{h}$} P809, 3 Quang Trung, {\fontencoding{T5}\selectfont H\h ai
Ch\^au, \DJ \`a N\~\abreve ng}, Vietnam}} }

\date{}

\maketitle


\begin{abstract}
We employ a set of recent, theoretically motivated, fits to non-perturbative unquenched gluon
propagators to check in how far double gluon exchange can be used to describe the soft sector of $pp$ scattering data (total and differential cross section). In
particular, we use the refined Gribov--Zwanziger gluon propagator (as arising from dealing with
the Gribov gauge fixing ambiguity) and the massive Cornwall-type gluon propagator (as motivated
from Dyson--Schwinger equations) in conjunction with a perturbative quark-gluon vertex, next to
a model based on the non-perturbative quark-gluon Maris--Tandy vertex, popular from Bethe--Salpeter descriptions of hadronic bound states. We compare the cross sections arising from these models with ``older'' ISR and more recent TOTEM and ATLAS data. The lower the value of total energy $\sqrt{s}$, the better the results appear to be.

\end{abstract}

\section{Introduction}
One of the most challenging problems in QCD (both from the phenomenological
and theoretical point of view) is to understand the hadronic spectrum and their interactions. More
specifically, we are interested in the
diffractive scattering process, which accounts for the exchange of Pomerons \cite{Pomeron}, in
the regime of low transferred momentum (the so-called \textit{soft Pomeron}).

Such process in the low transferred momentum regime is
still waiting for a first-principles explanation, since the behavior of its cross-section
observed at low transferred momentum suggests a violation of the Froissart--Martin bound
\cite{froissart,martin} (cf.~\cite{reviewPomeron1,Pomeronbook1,Pomeronbook2} for detailed reviews). The key observation is that at low transferred $\sqrt t$-momentum, the scattering amplitude does not decrease with $s$ as one would
expect on the basis of the
Froissart--Martin bound\footnote{Even if it is well-known that, at the scale of the soft Pomeron, the total cross section is
far from saturating the Froissart--Martin bound (see \cite{UniBound3} and references
therein), the puzzle is still there: why do other trajectories lead to falling cross sections, while the
soft Pomeron can lead to rising cross sections?} \cite{froissart,martin} (a modern and
interesting analysis can be found in \cite{UniBound3}).

In the nineteen-sixties, Chew and Frautschi \cite{Chew1961,Chew1962} observed
from plotting the spins of low-lying mesons (such as the $\rho$ meson) against their squared
masses, in a $t$-channel scattering process of hadrons, that they lie on straight lines.
This behavior was already predicted by Regge in 1959 in a paper \cite{regge} in which he studied
the analytic properties of the $S$-matrix as a function of angular momentum, when the latter is allowed to take any complex value.
In his theory, particles (low-lying mesons) exchanged during a scattering process of hadrons, such as $pp$ or $p\overline{p}$, fall along straight lines, the so-called  Regge trajectories,
\begin{eqnarray}
\alpha(t) ~=~ \alpha(0) + \alpha' t   \,,
\end{eqnarray}
either for positive or negative $t$. In the above equation, $\alpha$ is the Regge pole and $\alpha'$ stands for the slope of
the Pomeron trajectory, possibly dependent of $t$.

According to Regge theory, the Pomeron corresponds to the rightmost singularity in the complex
angular momentum plane, and by following the steps of \cite{Pomeronbook2} one should find
out that the expression of the amplitude $\mathcal{A}$ of a two-particle to two-particle scattering process
in the $t$-channel is proportional to $s^{\alpha(t)}$, with
\begin{eqnarray}
\label{amplt}
\mathcal{A} ~=~ i\beta(t) s^{\alpha(t)}\;.
\end{eqnarray}
In equation \eqref{amplt}, $\alpha(t)$ accounts for the leading Regge pole, and the residue
function $\beta(t)$ may be given by
\begin{eqnarray}
\beta(t) ~=~ A\exp{B t}\;.
\end{eqnarray}
Such a residue function accounts for the strength of the Pomeron coupling to protons.

A tricky point in the study of \textit{soft Pomerons} is that non-perturbative effects are expected
to play a key role, since the relevant region of phase space corresponds to the infrared regime: a discussion of Pomeron physics at low transferred momentum from first principles should be based on non-perturbative QCD. A deeper understanding of infrared soft Pomeron physics can teach us valuable lessons about non-perturbative QCD in return. A particular example of a source of non-perturbative effects (magnetic monopoles) was discussed by one of us in \cite{Canfora2016}.

In this paper we want to re-explore the relations between the Pomeron, from the
point of view of the Regge theory \cite{regge} (cf. \cite{reggebook1,reggebook2} for
detailed reviews) and functional non-perturbative approaches to QCD
$n$-point functions. In particular, within perturbative QCD, the Pomeron is usually modelled as the
exchange of
two (or even more pairs of) gluons (see \cite{Pomeronbook1,Pomeronbook2,Pomeronoriginal}).

However, the perturbative description of gluons leads to a Pomeron-singularity at $t=0$
(i.e. zero transferred momentum). Even though  including quarks in the proton
wave function can cancel the singularity, this procedure does not correctly reproduce the $t$-dependence  of the
differential cross section observed in the experiments \cite{Richards:1985re}. The singularity traces back to the massless pole in the gluon propagator at $k^2=0$ and it is commonly believed that non-perturbative corrections can give a softer behavior at small momenta $k$. For this reason, we will rely on an effective
non-perturbative gluon description based on the refined Gribov--Zwanziger framework, as well as massive Cornwall-type gluon propagator or even the inclusion of a non-perturbative Maris--Tandy vertex that is frequently used to model the quark-gluon interaction. These three descriptions include a modification of the standard $k^{2}$ behavior for the gluon propagator in the infrared regime.

The standard \textit{first principles} approach to the Pomeron based on QCD is the
so-called BFKL equation \cite{BFKL1,BFKL2,BFKL3}, which describes
the \textit{hard Pomeron} rather well (see \cite{Pomeronbook1,Pomeronbook2}). However, even within the BFKL equation approach the
issue related to the low $t$ behavior is still open. The present analysis
suggests again that the inclusion of the effects related to non-perturbative gluon effects in the BFKL formalism could be an important step to construct a
unified description of the hard and soft Pomeron.

The first analysis in this direction was in the paper of Landshoff and
Nachtmann \cite{LandshoffNachtmann} (see also \cite{NPPomeron1,NPPomeron2,NPPomeron3,NPPomeron4}). In these references,
the authors emphasized the fundamental role of the non-perturbative infrared
corrections to the gluon propagator in shedding light on the open issues in
Pomeron physics. In \cite{hancockross1,halzen,Henty:1995qr,Cudell:1993ui} the authors analysed the role of a non-perturbative gluon mass on the Pomeron in detail. In the present work we will continue this study, by incorporating the nowadays well-established lattice estimates of QCD two-point functions (unquenched case, that is, with dynamical quarks). We will also test a popular effective non-perturbative quark-gluon vertex that is supposed to describe the infrared interaction quite well, at least in obtaining the hadronic spectrum.

The paper is organized as follows. In Section \ref{sectionIII}, we will set the stage and plainly introduce the propagators we will use, next to a model inspired from the Bethe--Salpeter community that merges a non-perturbative gluon propagator and quark-gluon vertex into a single effective interaction. After that, we will briefly review how to describe the soft Pomeron as a two-gluon exchange following the existing literature, supplemented with more recent form factor fits that enter the computation \cite{Iachello:2004aq,Bijker:2004yu}. Section \ref{sectionIV} bundles our results and discussion thereof in comparison with different data sets.


\section{Soft Pomeron, propagators and cross sections} \label{sectionIII}

The simplest description of the Pomeron as the exchange between quarks of a pair
of gluons with the quantum numbers of the vacuum \cite{Pomeronoriginal} gives
the following amplitude\footnote{For comparison we use the same notation as \cite{halzen}.}:
\begin{equation}
{\mathcal{A}}=i\beta_{0}^{2}(\bar{u}\gamma_{\mu}u)(\bar{u}\gamma_{\mu}u)
\end{equation}
with $\beta_{0}$ the strength of the Pomeron coupling to quarks:
\begin{equation}
\beta_{0}^{2}=\frac{1}{36\pi^{2}}\int{d^{2}k\left[  g^{2}D(k^{2})\right]
^{2}}%
\end{equation}
As we already reminded, from the point of view of Pomeron physics,
one of the most common technical problems is the infrared divergence of the
perturbative gluon propagator with its massless pole behaviour, which often raises regularization issues (see
for instance \cite{Pomeronbook2}). A possible way out is to give a mass to the gluons.
Excluding a bare gluon mass due to its problems with unitarity, Cornwall proposed \cite{cornwall}
a mechanism, via solving a particular type of Dyson--Schwinger functional equations of motion, to generate the gluon mass dynamically and suggested a possible form for the propagator. This approach more recently inspired a whole series of papers gradually improving on his seminal work, leading to a satisfactory description. A selection of alternative relevant Dyson--Schwinger sources, next to other recent analytical takes at ``massive gluon propagators'' is \cite{Aguilar:2015bud,Dudal:2008}.

In particular, the Cornwall type of gluon propagator, tacitly assuming Landau gauge fixing $\partial_\mu A_\mu=0$, can be succinctly written as
\begin{eqnarray}\label{d1}
D_\text{Cornwall}(k)&=&\frac{1}{k^2+M^2(k^2)}\frac{33-2N_f}{18\pi^2}\frac{1}{4\pi \alpha_s}\ln\frac{k^2+4M^2(k^2)}{\Lambda^2}\\
M^2(k^2)&=& m^2\left(\frac{\ln\frac{k^2+4m^2}{\Lambda^2}}{\ln\frac{4m^2}{\Lambda^2}}\right)^{-\frac{12}{11}}
\end{eqnarray}
Notice that the Cornwall propagator solution\footnote{Or better said: fit to the numerical solution of the underlying Dyson--Schwinger equation. See \cite{Aguilar:2015bud} for references and similar fits to \eqref{d1}.} has a built-in (frozen) strong coupling constant factor $\alpha_s$. As parameters, we select $m=0.35~ \text{GeV}$ and $\Lambda=0.3~\text{GeV}$, while the number of active flavours is set equal to $4$.

In order to also get genuine access to non-perturbative gluon lattice data, we will also use the so-called Refined Gribov--Zwanziger (RGZ, \cite{Dudal:2008}) propagator,
\begin{eqnarray}\label{d2}
D_\text{RGZ}(k)&=&Z\frac{k^2+M^2}{k^4+(M^2+m^2)k^2+\lambda^4}
\end{eqnarray}
This propagator arises from considering the Gribov gauge-fixing ambiguity in any covariant gauge \cite{singer}. In particular in the Landau gauge, $\partial_\mu A_\mu=0$ allows for multiple gauge equivalent solutions \cite{Gri78,Zwanziger:1989mf,VZ}. The propagator \eqref{d2} arises from restricting the domain of the path integration to a smaller domain that counts less Gribov copies (more precisely, to a region free of infinitesimally connected gauge copies). The parameters $M^2$, $m^2$ and $\lambda^4$ are dynamical mass scales proportional to the fundamental QCD scale $\Lambda_{\text{QCD}}$ that come out of the (Refined) Gribov--Zwanziger formalism \cite{Dudal:2008}. $Z=Z(\mu)$ is a renormalization factor, suitably chosen so that the propagator is renormalized according to the MOM scheme, defined via
\begin{equation}\label{mom}
  D(k^2=\mu^2)=\frac{1}{\mu^2} \;,
\end{equation}
which is a scheme that can also be implemented non-perturbatively on the lattice.

In the absence of quarks, it was discussed in \cite{Cucchieri:2011ig,Oliveira:2012eh,DOV,DOR2012} that the functional form provides excellent fits to SU(2) and SU(3) lattice data. In \cite{BRR2014}, the fit was generalized to the case with $N_f$ dynamical quark flavours. It was argued, using $N_f=2+1+1$ flavours with realistic masses, that the gluon propagator is dominated by the 3 light flavours, and that one can set $N_f=3$ in the fitting parameters. For clarity, let us follow the same notation as in \cite{BRR2014}, where \eqref{d2} was rewritten as
\begin{equation}\label{BRR1}
  D(k)=Z\frac{k^2(k^2+M^2)}{k^4+k^2(M^2-\frac{13}{24}g^2\braket{A^2})+M^2m_0^2}
\end{equation}
with
\begin{eqnarray}
  m_0(N_f) &=& m_0(0)e^{AN_f}  \\
  g^2\braket{A^2}(N_f) &=& g^2\braket{A^2}(0)e^{-BN_f}
\end{eqnarray}
where $m_0(0) = 0.333~\text{GeV}$, $g^2\braket{A^2}(0)=7.856~\text{GeV}^2$, $A = 0.083$ and $B = 0.080$ and as said, $N_f=3$. The fits were obtained at $\mu=4.3~\text{GeV}$.

\begin{figure}[H]
   \centering
   \includegraphics[width = 0.7\textwidth]{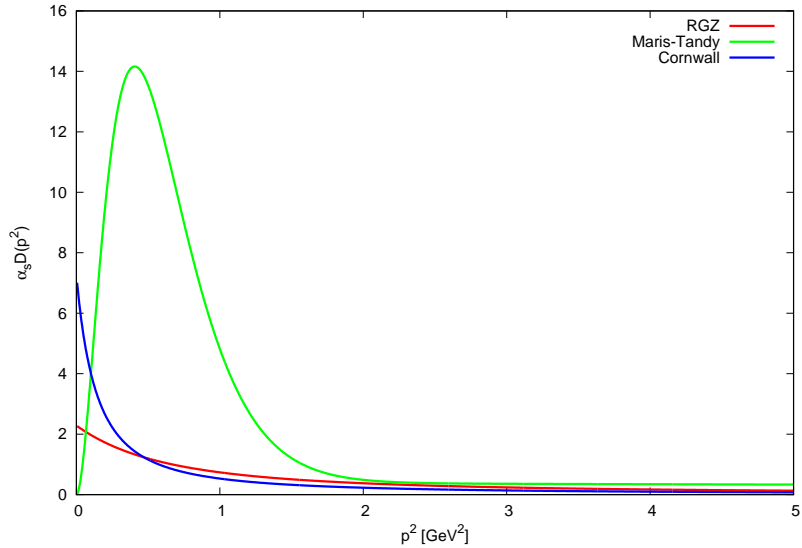}
   \caption{Comparison of different ``propagators~$\times$~coupling'' with infrared modifications. We can see clearly the propagator of the Maris--Tandy model is out of scale with respect to the RGZ and Cornwall ones.}\label{fig_propagators}
\end{figure}
Since we are eventually considering the (double) gluon exchange between the constituent quarks of the protons, the QCD coupling between each gluon to the two quarks inside the protons will happen via the quark-gluon vertex, giving the combination $g^2D(k^2)$ as relevant player. For that reason, in Figure \ref{fig_propagators} we have shown this combination for both propagators \eqref{d1} and \eqref{d2}. The Figure also contains a popular modelling of the vertex, proposed by Maris--Tandy in \cite{Maris:1999nt}, see also \cite{rojas} and the recent review \cite{Eichmann:2016yit}. This last one has provided rather sensible results in the construction of meson and baryon properties from a functional viewpoint. In our notation, it is given by
\begin{eqnarray}\label{d3}
\alpha_s D_\text{Maris--Tandy}(k)&=&\pi \eta^7\frac{k^4}{\Lambda^4}e^{-\eta^2\frac{k^2}{\Lambda^2}}+\frac{2\pi\gamma_m\left(1-e^{-\frac{k^2}{\Lambda_t^2}}\right)}{\ln\left(e^2-1+\left(1+\frac{k^2}{\Lambda_{\text{QCD}}^2}\right)^2\right)}
\end{eqnarray}
where $\alpha_{s}=g^{2}/4\pi$ is the strong coupling constant, while \cite{Eichmann:2016yit} $\gamma_m=\frac{12}{25}$ for $4$ active flavours, $\Lambda_{\text{QCD}}=0.234~\text{GeV}$, $\Lambda=0.72~\text{GeV}$, $\Lambda_t=1~\text{GeV}$ and the parameter $\eta$ is usually varied between $1.6$ and $2$, without affecting the probed meson or baryon properties too much. Our graph corresponds to $\eta=2$. The Maris--Tandy modelling tacitly assumes that the quark-gluon vertex remains proportional to its bare Lorentz form, $\gamma_\mu$, which evidently does not need to be the case. However, it is a welcome feature here since the derivation of the amplitudes with double gluon exchange was explicitly based on a vertex $\propto \gamma_\mu$.

Indeed, returning to the Pomeron contribution, the amplitude for the elastic proton-proton scattering with two-gluon exchange
can be written (as worked out in full detail in \cite{Cudell:1993ui}, thereby correcting an error that originated in foregoing works \cite{halzen,Cudell:1991uz}) as\footnote{The $\alpha_{s}^{2}$ prefactor usually present in \eqref{ast} has been omitted here, as we already introduced it via the propagators.}
\begin{equation}\label{ast}
\mathcal{A}(s,t)=is8\alpha_s^2\left[  T_{1}-T_{2}\right]
\end{equation}
with
\begin{align}
T_{1}  &  =\int_{0}^{s}{d^{2}kD\left(  k+\frac{q}{2}\right)  D\left(
-k+\frac{q}{2}\right)  \left[  G_{p}\left(  q,0\right)  \right]  ^{2}%
}\label{T1}\\
T_{2}  &  =\int_{0}^{s}{d^{2}kD\left(  k+\frac{q}{2}\right)  D\left(
-k+\frac{q}{2}\right)  G_{p}\left(  q,k-\frac{q}{2}\right)  \left[
2G_{p}\left(  q,0\right)  -G_{p}\left(  q,k-\frac{q}{2}\right)  \right]  }
\label{T2}%
\end{align}
where $D(k)$ is the gluon propagator, and $G_{p}(q,k)$ a convolution of the proton wave
function. In the following we will use the parametrization given by
\cite{Cudell:1993ui} (here $t=-q^{2}\leq0$, with $f=1$ in equations \eqref{G_functions} of \cite{Cudell:1993ui})
\begin{eqnarray}\label{G_functions}
G_{p}(q,k)&=&F_{1}(t)\;, \nonumber\\
G_{p}\left(q,k-\frac{q}{2}\right)&=&F_{1}\left(-3k^2  +\frac{t}{4}\right) \;.
\end{eqnarray}
$F_{1}(t)$ is the Dirac form factor of the proton, which in older works reads
\[
F_{1}(t)=\frac{4m_{p}^{2}-2.79t}{4m_{p}^{2}-t}\frac{1}{\left(  1-\frac
{t}{0.71}\right)  ^{2}}%
\]
but we will update it with
\begin{eqnarray}\label{Iachelo_form_factors}
F_1(t)&=&\frac{1}{2}g(-t)\left[1-\beta_\omega-\beta_\phi+\frac{\beta_\omega m_\omega^2}{m_\omega^2-t}+\frac{\beta_\phi m_\phi^2}{m_\phi^2-t}\right]\nonumber\\
&&+\frac{1}{2}g(-t)\left[1-\beta_\rho+\beta_\rho\frac{m_\rho^2+\frac{8}{\pi}\Gamma_\rho m_\pi}{m_\rho^2-t+(4m_\pi^2-t)\Gamma_\rho\frac{\alpha(-t)}{m_\pi}}\right] \;,\\
g(-t)&=&\frac{1}{(1-\gamma t)^2}\;, \nonumber\\
\alpha(-t)&=&\frac{2}{\pi}\sqrt\frac{4m_\pi^2-t}{-t}\ln\frac{\sqrt{-t+4m_\pi^2}+\sqrt{-t}}{2m_\pi}\;,\nonumber
\end{eqnarray}
as developed in \cite{Iachello:2004aq,Bijker:2004yu}. We have used $\beta_\rho=0.672$, $\beta_\omega=1.102$, $\beta_\phi=0.112$, $\gamma=0.25~\text{GeV}^{-2}$, $m_\pi=0.135~\text{GeV}$ $m_\rho=0.775~\text{GeV}$, $m_\omega=0.782~\text{GeV}$, $m_\phi=1.019~\text{GeV}$ and $\Gamma_\rho=0.112~\text{GeV}$. It is worth to emphasize that the results obtained with the form factors \cite{Iachello:2004aq,Bijker:2004yu} in equations \eqref{Iachelo_form_factors}, are closer to the experimental data than the results obtained with the old form ones in \cite{halzen,Cudell:1993ui}.

The properties of interest of the soft Pomeron can then be extracted from numerically evaluating the above integrals using polar coordinates. From the scattering amplitude, one gets
\begin{eqnarray}\label{ast2}
  \sigma &=& \frac{A(s,0)}{is}
\end{eqnarray}
for the total cross section, and
\begin{eqnarray}
  \frac{d\sigma}{dt} &=& \frac{|A(s,t)|^2}{16\pi^2 s}
\end{eqnarray}
for the differential cross section. Both quantities are in principle connected trough the optical theorem.

A small note, to accommodate for the energy-dependent part of the (differential) cross section, we actually have shown
\begin{eqnarray}
  \frac{d\sigma}{dt} &=& s^{0.168}\frac{|A(s,t)|^2}{16\pi^2 s} \;,
\end{eqnarray}
see \cite{halzen}. This added prefactor $s^{0.168}$ is necessary to improve the fit, and it also accounts for the energy $s$-dependence in the total cross section, which was estimated to behave as \cite{Pomeronbook1}
\begin{equation}
  \sigma\sim 22.7 s^{0.08} \;.
\end{equation}
Indeed, looking at \eqref{ast2} in combination with \eqref{ast}, it is immediately realized that this results in an $s$-independent (constant) $\sigma$. The two-gluon exchange picture of the Pomeron thus ought to solely describe the $s\to0$ very soft part of the (total) cross section, the rest would come from neglected higher order exchange diagrams.

\section{Results and comparison with experimental data} \label{sectionIV}

\begin{figure}[ptb]
\subfloat[][]{\includegraphics[width=.50\textwidth]{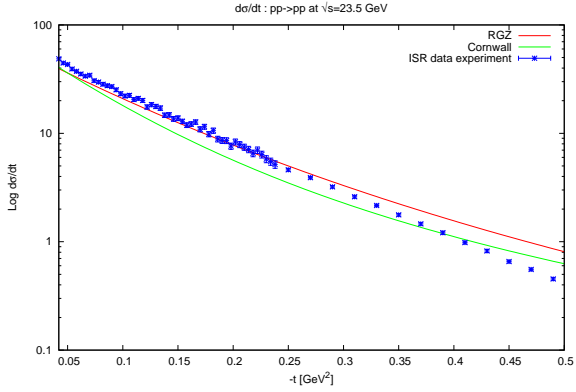}}
\subfloat[][]{\includegraphics[width=.50\textwidth]{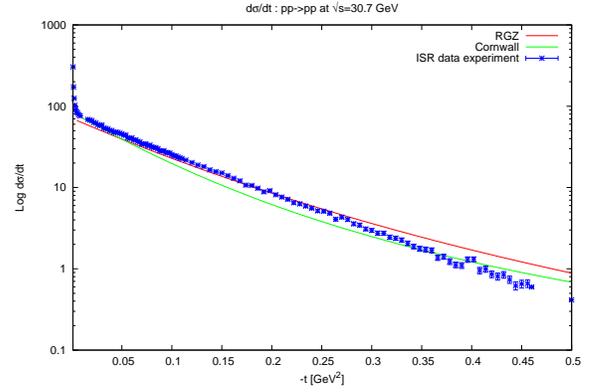}} \ \centering
\subfloat[][]{\includegraphics[width=.50\textwidth]{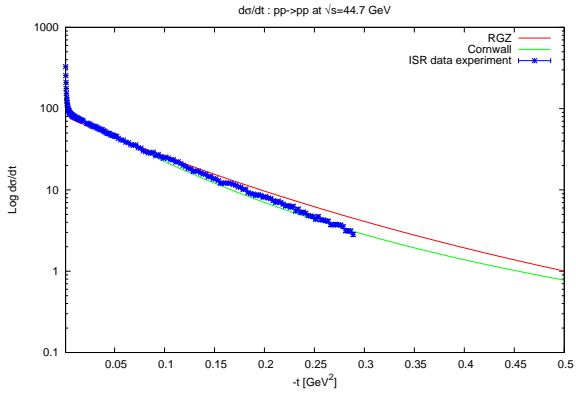}}
\subfloat[][]{\includegraphics[width=.50\textwidth]{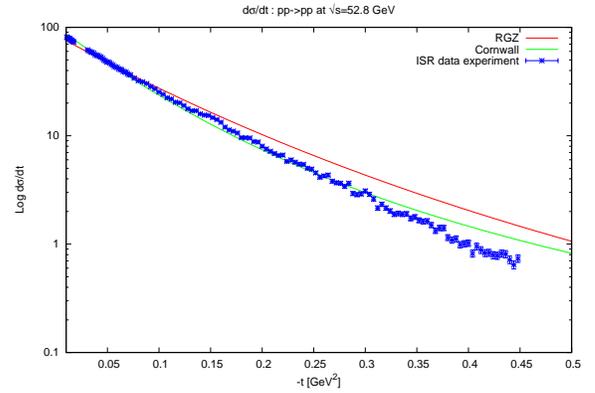}} \ \centering
\subfloat[][]{\includegraphics[width=.50\textwidth]{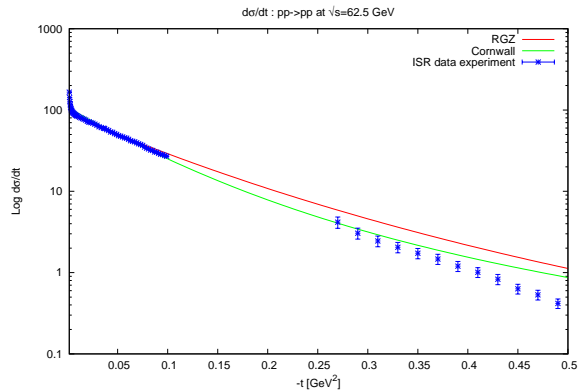}}
\caption{Comparison among RGZ (red line), Cornwall
(green line) and ISR data values (blue line) at (a) $\sqrt{s}=23.5$ GeV (blue line); (b) $\sqrt{s}=30.7$ GeV (blue line); (c) $\sqrt{s}=44.7$ GeV; (d) $\sqrt{s}=52.8$ GeV; (e) $\sqrt{s}=62.5$ GeV.}\label{ISR}
\end{figure}

In Figure \ref{ISR} we have compared the differential cross section for proton-proton scattering between the theoretical RGZ and Cornwall prediction versus the ISR data at $\sqrt{s}=23.5$ GeV (a), $\sqrt{s}=30.7$ GeV (b), $\sqrt{s}=44.7$ GeV (c), $\sqrt{s}=52.8$ GeV (d), $\sqrt{s}=62.5$ GeV. This data was collected from \cite{Breakstone:1984te}. We have also worked with the Maris--Tandy parametrization, but this is not shown on the plot, as the outcome is a few orders of magnitude too large. This might suggest that a fully non-perturbative quark-gluon vertex, incorporating more general tensor structures than $\gamma_\mu$, could be rather relevant in the Pomeron sector. It would be interesting to investigate this further, using more elaborate vertex ans\"atze, see for example \cite{Williams:2014iea,Binosi:2016wcx}. In the RGZ case, the coupling constant $\alpha_s$ is similar as the one used in \cite{Cudell:1993ui}, left as a free parameter to be fitted. We have found that freezing the value to $\alpha_s=0.35$ gives satisfactory results for all ISR data sets.

We have also considered the case of an effective (infrared and ultraviolet) coupling constant $\alpha_s(k)$ running with momentum $k$ when coupled to a gluon propagator $D(k)$. Using the proposed form of \cite{Deur:2016opc} as input for $\alpha(k)$, the result for the differential cross section was an order of magnitude too large. This being said, since we used the RGZ gluon propagator renormalized at $\mu=4.3~\text{GeV}$, the natural choice might be to consider $\alpha_s(\mu=4.3~\text{GeV})$. A rough estimate based on Fig.~1 of \cite{Deur:2016opc} learns that our fit $\alpha_s=0.35$ is about a factor of 2 larger than the experimental value of $\alpha_s(4.3)$. The fact that we need a somewhat larger value is not unexpected, since at $4.3~\text{GeV}$ we are almost in the perturbative region, while we are after all describing soft, infrared physics where the coupling is somewhat larger, as can also be seen from Fig.1~of \cite{Deur:2016opc}.

We can see the theoretical descriptions do not fit very well at larger $t$-momenta. In \cite{Cudell:1993ui}, this is attributed to the effects of the odderon that become dominant at large $t$. More generally, the mismatch at large $t$ is caused by a lack of multiple ($>2$) gluon exchanges.

\begin{figure}[ptb]
\centering
\subfloat[][]{\includegraphics[width=.50\textwidth]{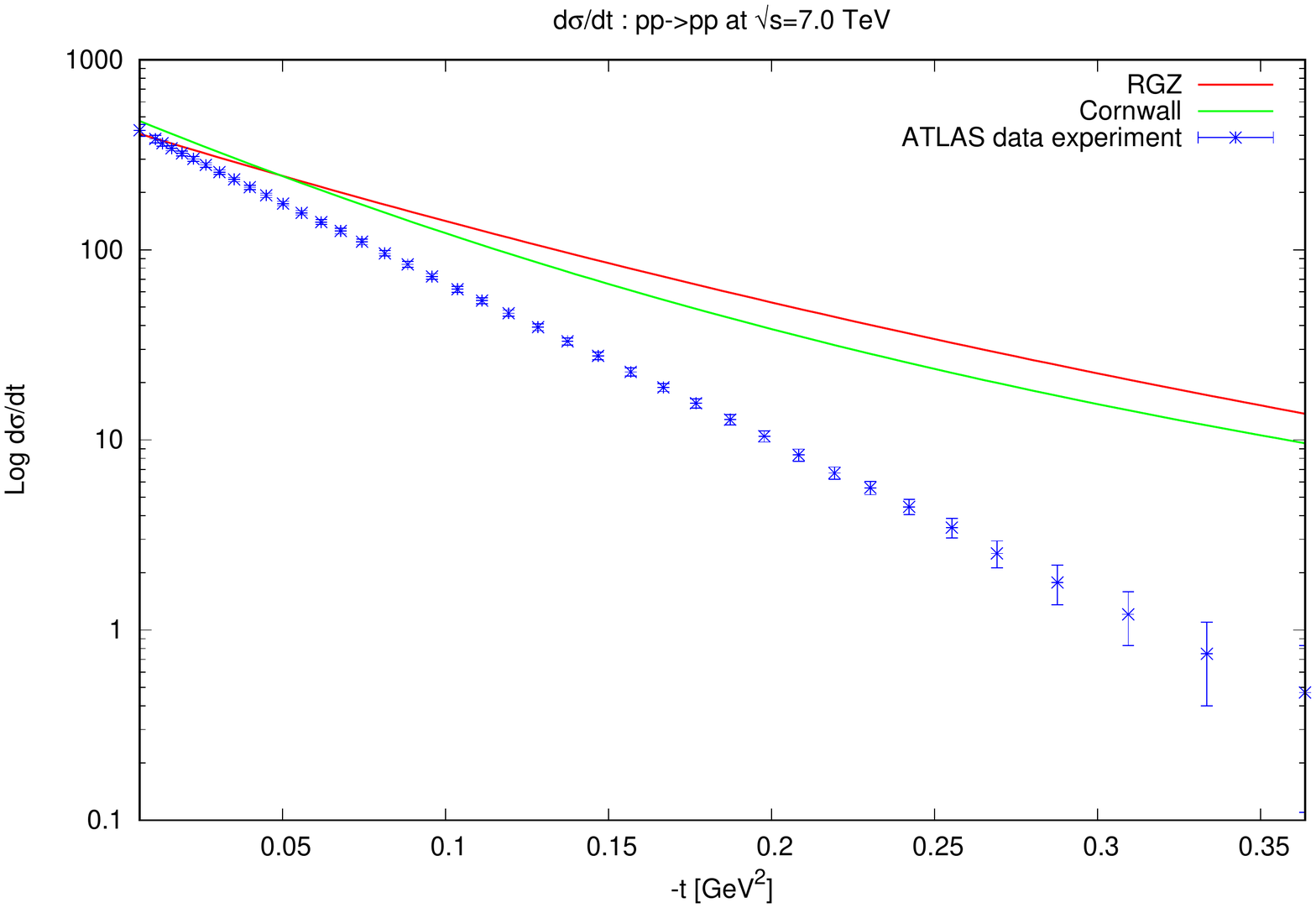}}
\subfloat[][]{\includegraphics[width=.50\textwidth]{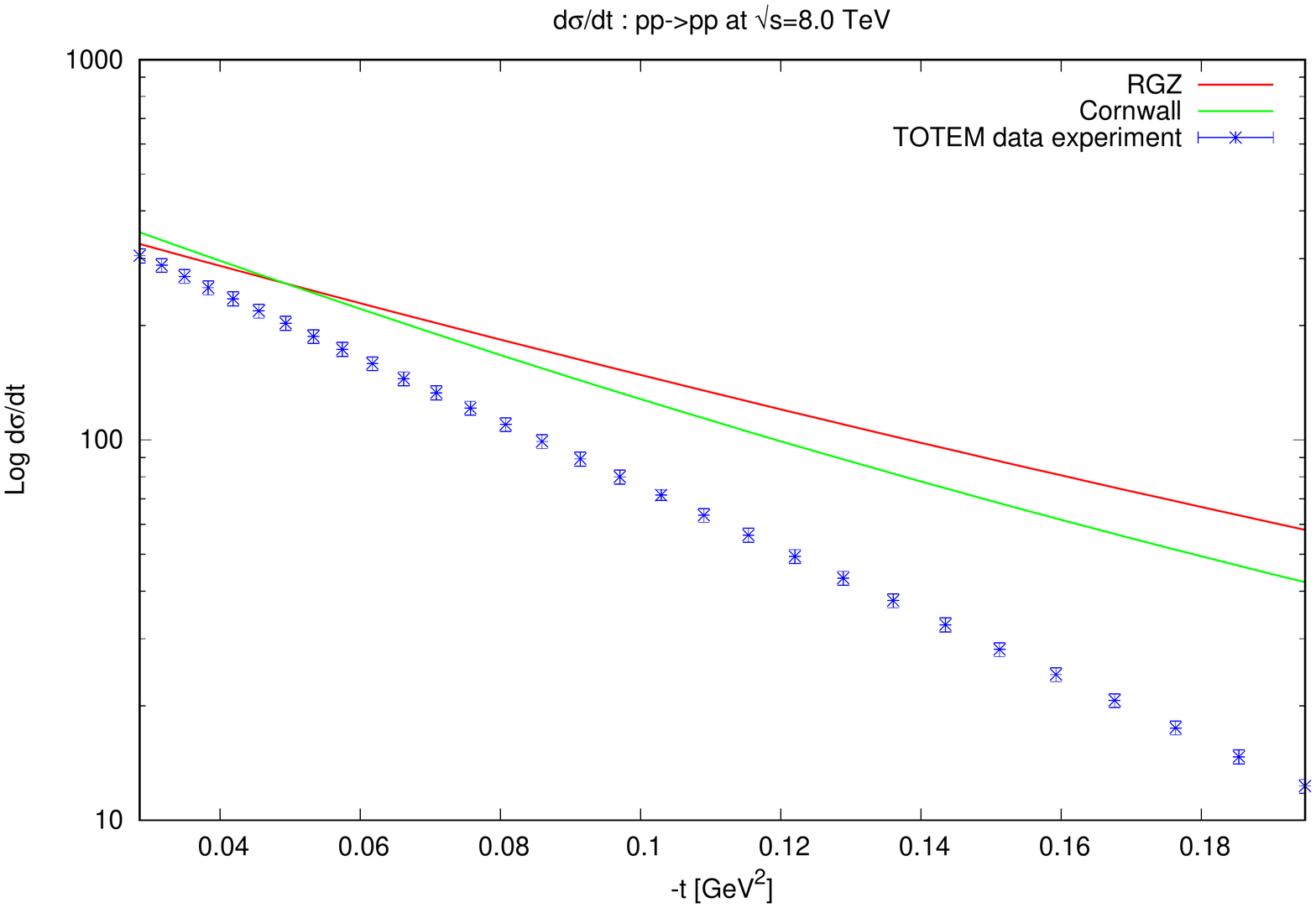}}
\caption{Comparison among RGZ (red line), Cornwall
(green line) and experimental data values (blue line) of (a) ATLAS at $\sqrt{s}=7.0$ TeV ; (b) TOTEM at $\sqrt{s}=8.0$ TeV. We can appreciate here that our theoretical description fails because we are only taking two gluon-exchange into account.}\label{TOTEMATLAS}
\end{figure}

In Figure \ref{TOTEMATLAS} we show the differential cross section for RGZ, Cornwall and ATLAS data at
$\sqrt{s}=7$ TeV, while the TOTEM data are at $\sqrt{s}=8$ TeV. The TOTEM data were gathered from \cite{totem} and the ATLAS data from \cite{atlas}.

We can see the disagreement with respect to the experimental data starting
from these values of $\sqrt{s}$. Of course, this was quite expected since at such
large $s$, the hard Pomeron (well-described by the BKFL equation) becomes
dominant \cite{froissart,Pomeronbook1}. In other words, the principled way to
push our current description to higher values of $s$ would be to solve the full BFKL
equation using a non-perturbative propagator, instead of the perturbative one usually employed in BFKL, as basic ingredient. We hope to come back to this interesting issue, which
unfortunately presents quite remarkable technical difficulties at the numerical level, in future work.

\subsection{Exponential fit}

As mentioned before, the scattering amplitude may be written as
\begin{eqnarray}
\mathcal{A} ~=~ i\beta(t) s^{\alpha(t)}\;,
\end{eqnarray}
where the residue function $\beta(t)$ is given by
\begin{eqnarray}
\beta(t) ~=~ A\exp{B t}\;.
\end{eqnarray}
Such a residue function accounts for the strength of the Pomeron coupling to
protons. Since the differential elastic cross section is given by
\begin{eqnarray}
\frac{d \sigma}{dt} ~\propto~ \vert \mathcal{A} \vert^{2}\;,
\end{eqnarray}
the slope parameter of the Regge trajectory can be derived from
\begin{eqnarray}
\label{slope}
b(s,t) ~=~ \frac{d}{dt} \left[ \ln \left( \frac{d\sigma}{dt} \right) \right]\;,
\end{eqnarray}
which gives
\begin{eqnarray}
\label{exp.slope}
b(s,t) ~=~ 2B + 2\alpha'(t)\ln s\;.
\end{eqnarray}
In the above equation, \eqref{exp.slope}, $\alpha'(t)$ stands for the Pomeron trajectory slope,
$\alpha(t) = \alpha_{0} + \alpha'(t)t$,
and is, in principle, $t$-dependent. However, for sufficiently small values of $\vert t \vert$
one may consider the Pomeron trajectory as being linear, which yields a $t$-independent slope
$\alpha'$. Within this regime (and for certain values of $s$) the differential elastic cross
section may be suitably fitted by
\begin{eqnarray}
\label{linearfit}
\frac{d \sigma}{dt} ~=~ ae^{bt}\;.
\end{eqnarray}
Though, according to the CERN ISR data,
\cite{Breakstone:1984te}, for $\sqrt{s} = \unit{31}{\giga\electronvolt}$ the data shows a better fit with a quadratic
dependence of $\vert t \vert$ in the exponential, such as
\begin{eqnarray}
\frac{\text{d}\sigma}{\text{d} t} ~=~ a\exp\left\{  Bt + Ct^{2} \right\}\;.
\end{eqnarray}
In this case the slope parameter is given as a function of $\vert t \vert$, by $b = B
+C\vert t \vert$. Moreover, also recent data indicate that $b$ is not a constant but a
function of $\vert t \vert$, which means that the Pomeron trajectory is not to be taken as
linear but rather as non-linear, and the slope parameter should be described by
equation \eqref{slope}, see for instance \cite{totem}. At the same time, for the energy regimes $\sqrt{s} =
\unit{53}{\giga\electronvolt}$ and $\sqrt{s} = \unit{62}{\giga\electronvolt}$, the simplest
exponential fit of the differential cross section \eqref{linearfit} just works out fine.

In our work this simple exponential fit will be used, since we shall remain within the above
energy level $\sqrt{s} = \unit{31}{\giga\electronvolt}$,
$\sqrt{s} = \unit{53}{\giga\electronvolt}$ and $\sqrt{s} = \unit{62}{\giga\electronvolt}$, with
considerably small values of $\vert t \vert$. In this section the values of the slope
parameter $b$ and of the nuclear intersect parameter $a$, obtained within the RGZ framework, will be presented\footnote{The results from the Cornwall propagator input are very similar.}. Furthermore, by comparing with
results from experimental data, \cite{Breakstone:1984te} and \cite{Ambrosio:1982zj}, it will
be clear that this is effectively the regime where our analysed model works out best.

The range of values of the squared
four-momentum transfer parameter $t$ and of the kinetic energy $\sqrt{s}$ were chosen in order
to compare with equivalent results obtained in \cite{Breakstone:1984te} and
\cite{Ambrosio:1982zj}.

According to \cite{Breakstone:1984te}, for $\sqrt{s} =
\unit{31}{\giga\electronvolt}$ and $\unit{0.05}{\giga\electronvolt\squared} < -t < \unit{0.15}{\giga\electronvolt\squared}$ they have found that
$a_{\text{Breakstone}} = \unit{93.0 \pm 5.5}{mb/\giga\electronvolt\squared}$ and $b_{\text{Breakstone}} =
\unit{11.70 \pm 0.62}{\giga\electronvolt^{-2}}$, for $pp$ scattering. From the first line of
Table \ref{rgzslope} it can be seen that, for this same range of values, we get $a =
\unit{91.53}{mb/\giga\electronvolt\squared}$ and $b = \unit{11.87}{\giga\electronvolt^{-2}}$,
which clearly agree well with \cite{Breakstone:1984te}.

\begin{table}[h]
\begin{center}
\begin{tabular}{|c|c|c|c|}
\hline
$\unit{\sqrt{s\;\;}}{(\giga\electronvolt)}$ & $\unit{-t\;\;}{(\giga\electronvolt\squared)}$ &
$\unit{a\;\;}{(mb/\giga\electronvolt^{2})}$ & $\unit{b\;\;}{(\giga\electronvolt^{-2})}$
\\
\hline
$31$ & $0.05$ -- $0.15$ & $91.53$ & $11.87$
\\
\hline
$31$ & $0.17$ -- $0.85$ & $56.62$ & $9.05$
\\
\hline
$52.8$ & $10^{-7}$ -- $0.05$ & $116.35$ & $13.02$
\\
\hline
$52.8$ & $0.09$ -- $1$ & $92.45$ & $10.42$
\\
\hline
\end{tabular}
\caption{Nuclear intersection $a$ and slope parameter $b$ obtained from the RGZ framework applied to the proton-proton scattering process. In order to obtain the values of this table, we made use of the data of
\cite{Bijker:2004yu}.}\label{rgzslope}
\end{center}
\end{table}

Still looking at Breakstone \emph{et al.}'s results, for $\sqrt{s} = \unit{31}{\giga\electronvolt}$
and $\unit{0.17}{\giga\electronvolt\squared} < -t < \unit{0.85}{\giga\electronvolt\squared}$
they have $a_{\text{Breakstone}} = \unit{74.0 \pm 3.6}{mb/\giga\electronvolt\squared}$ and
$b_{\text{Breakstone}} = \unit{10.92 \pm 0.15}{\giga\electronvolt^{-2}}$.
For this same range of values one may clearly see, from the second
line of Table \ref{rgzslope}, that both our results for $a$ and $b$ do not agree with those
of \cite{Breakstone:1984te}. Furthermore, note that our nuclear intersect parameter $a$ differs strongly
from the one found by Breakstone \emph{et al.} Such disagreement may be due to the larger absolute
value of $t$ considered in this case.

Now, let us take a look at Ambrosio \emph{et al.}'s results, of \cite{Ambrosio:1982zj}. In
the referred work they analyse the energy level
$\sqrt{s} = \unit{52.8}{\giga\electronvolt}$ for $t$ within the range
$-t < \unit{0.05}{\giga\electronvolt\squared}$ and
$\unit{0.09}{\giga\electronvolt\squared} <  -t <
\unit{1}{\giga\electronvolt\squared}$. In order to obtain something comparable to their result
within the first range of values, we studied the range
$\unit{10^{-7}}{\giga\electronvolt\squared} <  -t <
\unit{0.05}{\giga\electronvolt\squared}$, which can be seen from the third line of Table
\ref{rgzslope}. Ambrosio \emph{et al.} found that $a_{\text{Ambrosio}} = \unit{96.6 \pm
1.9}{mb/\giga\electronvolt\squared}$, while we get $a =
\unit{116.35}{mb/\giga\electronvolt\squared}$, which is a bit greater than their result. Still
within this range of values, they obtained a slope parameter of $b_{\text{Ambrosio}} = \unit{13.09
\pm 0.58}{\giga\electronvolt^{-2}}$, and we find $b =
\unit{13.02}{\giga\electronvolt^{-2}}$, which is in good agreement with them.

Now, still considering $\sqrt{s} = \unit{52.8}{\giga\electronvolt}$ but with
$\unit{0.09}{\giga\electronvolt\squared} <  -t <
\unit{1}{\giga\electronvolt\squared}$, they obtained $b_{\text{Ambrosio}} = \unit{10.34
\pm 0.25}{\giga\electronvolt^{-2}}$, while we have $b = \unit{10.42}{\giga\electronvolt^{-2}}$.
Our nuclear intersect parameter is $a = \unit{92.45}{mb/\giga\electronvolt\squared}$, which is
$a_{\text{Ambrosio}} = \unit{96.6 \pm 1.9}{mb/\giga\electronvolt\squared}$, which
is not so far from our result.

\section{Conclusions and outlook} \label{sectionIV}

In this work, we have updated the description of the soft Pomeron as the exchange of two
non-perturbative gluons. We have used well-motivated theoretical descriptions of the non-perturbative gluon propagator that are matchable with state-of-the art lattice data.
In particular we compare the RGZ, massive Cornwall-type gluon propagators, in addition to a Maris--Tandy effective quark-gluon vertex that incorporates the gluon propagator. The last model appears to be out of scale with respect to the experimental data, at least from the Pomeron point of view. It is interesting and yet unsatisfactory that the success of the latter vertex model in the construction of hadronic bound states via the Bethe--Salpeter equation formalism cannot be extended to yet another intrinsically non-perturbative QCD sector (soft scattering), even though exactly the same quark-gluon interactions are the essential ingredients.

The total cross section for the $pp\rightarrow pp$ process for the other two models presents a reasonable fit with ISR data --- i.e.~at relatively low total energy $\sqrt{s}$. Concerning larger total energies $\sqrt{s}$, the total cross sections obtained from the theoretical propagators start to depart from the more recent TOTEM and ATLAS data. This result is somehow expected, as the larger the values of $\sqrt{s}$, the larger the contribution of the hard Pomeron \cite{Pomeronbook2}. From this perspective, a possible interesting future work would be to analyse the modified BFKL equation in
which the perturbative gluon propagator is replaced by a non-perturbative and lattice-motivated one, to shed some light on to the, until now, blurred link between soft and hard Pomeron physics. The idea to analyze the BFKL equation with a modified gluon propagator is not new (see, for instance, \cite{Levin:2014bwa} and references therein). On the other hand, the BFKL equation with the RGZ propagator has not been analyzed yet and it is an intriguing topic to which we hope to come back to in a future publication.
\section*{Acknowledgements}
We wish to thank J.-R.~Cudell, F.~Iachello, G.~Krein, A.~Natale and J.~Rodriguez-Quintero for helpful communication during the preparation of this work.
 F.~C. was partially supported by Fondecyt Grant 1160137.  I.~F.~J. is grateful for the hospitality at UERJ (Brazil) and UGent (Belgium) during the initial stages of this research. P.~P. was partially supported from Fondecyt grant 1140155 and also thanks the Faculty of Mathematics and Physics of Charles University in Prague, Czech Republic for the kind hospitality at different stages of this work.
 P.~S-R is supported by Fondecyt grant 3160581. The Centro de Estudios Cient\'{\i}ficos (CECS) is funded by the Chilean Government through the Centers of Excellence Base Financing Program of Conicyt.

\end{document}